\documentclass[conference]{IEEEtran}
\IEEEoverridecommandlockouts
\usepackage{cite}
\usepackage{amsmath,amssymb,amsfonts}
\usepackage{graphicx}
\usepackage{textcomp}
\usepackage{xcolor}
\definecolor{mygray}{rgb}{0.5,0.5,0.5}
\definecolor{changed}{rgb}{0,0,0}
\definecolor{changed2}{rgb}{0.0,0,0}
\definecolor{myblue}{rgb}{0,0,.6}
\usepackage{textpos}
\usepackage{epstopdf}
\def\BibTeX{{\rm B\kern-.05em{\sc i\kern-.025em b}\kern-.08em
    T\kern-.1667em\lower.7ex\hbox{E}\kern-.125emX}}
\usepackage[scaled=0.92]{helvet}    

\usepackage{hyperref}
\hypersetup{
    pdftitle={Round-Trip Energy Efficiency and Energy-Efficiency Fade Estimation for Battery Passport},    
    pdfauthor={Camiel Beckers, Erik Hoedemaekers, Arda Dagkilic and Henk Jan Bergveld},     
    pdfkeywords={Battery Passport} {Electric Vehicle} {Battery Efficiency} {Energy Efficiency} {Multiple Linear Regression}
}

\usepackage{eso-pic}
\AddToShipoutPictureBG{%
  \AtTextUpperLeft{%
    \put(\LenToUnit{0cm},\LenToUnit{.5cm}){%
        \begin{minipage}[t]{24cm}
          \small
            Accepted author version of {\color{myblue}\href{https://doi.org/10.1109/VPPC60535.2023.10403325}{10.1109/VPPC60535.2023.10403325}}.
          \end{minipage}
    }%
  }%
}

\begin{document}
\IEEEpubid{\begin{minipage}{\textwidth}\ \\[50pt] \centering
  \copyright 2023 IEEE.  Personal use of this material is permitted.  Permission from IEEE must be obtained for all other uses, in any current or future media, including reprinting/republishing this material for advertising or promotional purposes, creating new collective works, for resale or redistribution to servers or lists, or reuse of any copyrighted component of this work in other works.
\end{minipage}}


\title{Round-Trip Energy Efficiency and Energy-\linebreak Efficiency Fade Estimation for Battery Passport
\thanks{This work has received financial support from the Ministry of Economic Affairs and Climate, under the grant ‘R\&D Mobility Sectors’ carried out by the Netherlands Enterprise Agency.}
}

\author{\IEEEauthorblockN{Camiel Beckers\IEEEauthorrefmark{1}, Erik Hoedemaekers\IEEEauthorrefmark{1}, Arda Dagkilic\IEEEauthorrefmark{2} and Henk Jan Bergveld\IEEEauthorrefmark{3}\IEEEauthorrefmark{4}
\vspace{0.3cm}
}
\IEEEauthorblockA{\IEEEauthorrefmark{1}Powertrains Dept., TNO, Helmond, The Netherlands, Email: camiel.beckers@tno.nl
}
\IEEEauthorblockA{\IEEEauthorrefmark{2}VDL Enabling Transport Solutions, Valkenswaard, The Netherlands
}
\IEEEauthorblockA{\IEEEauthorrefmark{3}Dept. of Electrical Engineering, Eindhoven University of Technology, Eindhoven, The Netherlands
}
\IEEEauthorblockA{\IEEEauthorrefmark{4}NXP Semiconductors, Eindhoven, The Netherlands
}
}

\maketitle

\begin{abstract}
The battery passport is proposed as a method to make the use and remaining value of batteries more transparent.
The future EU Battery Directive requests this passport to contain the round-trip energy efficiency and its fade.
%
In this paper, an algorithm is presented and demonstrated that estimates the round-trip energy efficiency of a battery pack.
The algorithm identifies round trips based on battery current and SoC and characterizes these round trips based on certain conditions.
{\color{changed}2D efficiency maps are created as a function of the conditions `temperature' and `RMS C-rate'.}
{\color{changed}The maps are parameterized using multiple linear regression, which allows comparison of the efficiency under the same conditions.}
Analyzing data from three battery-electric buses over a period of 3.5\,years reveals an efficiency fade of up to 0.86\,percent point.
\end{abstract}

\section{Introduction}
In the transition to electrically powered road transport, most Electric Vehicles (EVs) rely on Lithium-Ion Batteries (LIBs).
In 2021, the total yearly EV battery request was near 350\,GWh, and this number is expected to increase to at least 2\,TWh by 2030 \cite{IEA2022b}.
Simultaneously, more EVs are entering the second-hand market and batteries of depreciated vehicles are repurposed in second-use applications such as stationary {\color{changed2} energy} storage.
This brings challenges concerning the re-use of LIBs from EVs, including the fact that reliable methods are required to grade the quality of the LIBs \cite{IEA2022a}.

Driven by the growing EV battery market, the European Commission proposed a new EU Battery Directive \cite{EuropeanUnion2023} in 2020, repealing Directive 2006/66/EC and amending Regulation (EU) No 2019/1020.
Part of this proposed directive is the introduction of a battery passport that gathers key information for every EV battery.
This information is supposed to be made accessible online and also includes health information about the EV battery, which is to be updated throughout the battery's lifetime.
This information will play a key role in determining the current value of the battery, e.g., in case of resale, its applicability in other applications, e.g., second-life applications in grid energy storage, and will help in defining the right moment to recycle a battery.
{\color{changed} The realization of the battery passport will pose several technical challenges;}
{\color{changed2} Firstly, research is required to translate the abstract legislation into practical measures and implications. Specifically, the battery-health parameters mentioned in the EU directive are to be defined from a technical perspective and appropriate algorithms should be developed to estimate these parameters based on real-world data.}
{\color{changed} Secondly, the {\color{changed2}resulting} information has to be stored in the passport permanently and reliably.

One of the parameters that is expected to be part of the EU-legislated battery passport is the round-trip energy efficiency $\eta_{RT,e}$ and the fade, or decrease over time, of $\eta_{RT,e}$ with respect to that of the new battery \cite{EuropeanUnion2023}.
The round-trip energy efficiency $\eta_{RT,e}$, also named electrical efficiency, quantifies the energy that can be withdrawn from a battery with respect to the energy that is required to charge the battery back to the same State of Charge ($SoC$) \cite{Eftekhari2017a}.
%
Thereby, $\eta_{RT,e}$ incorporates the effects of various battery parameters, e.g., the battery impedance, under real operating conditions and characterizes the battery losses in an interpretable way.

Energy inefficiency is caused by a combination of 1) the polarization voltage during (dis)charging and 2) the coulombic losses, which are often represented by the coulombic efficiency \cite{Zurfi2017}.
The coulombic efficiency of Li-ion cells is generally higher than 99\%, and requires precision lab equipment to be measured accurately \cite{Smith2010a}.
The energy efficiency is per definition lower than the coulombic efficiency and is typically in the order of 96\% {\color{changed2}for LIBs}, but can be as low as 85.5\% for specific situations \cite[Fig.~6]{Schimpe2018}.
As a battery ages, both the decrease of the coulombic efficiency and the increase in the internal impedance, caused by SEI layer growth and lithium plating, will decrease $\eta_{RT,e}$ \cite{Ahmadi2014}.
Furthermore, the energy efficiency of a cell is reported to depend on $SoC$ \cite{Liu2017}, Depth of Discharge ($DoD$) \cite{Liu2017}, C-rate \cite{Kang2014}, and battery temperature $T$ \cite{Xu2018}.

Round-trip efficiency measurements often involve prescribing predetermined discharge/charge cycles to a battery or cell \cite{USDepartmentOfEnergy2001a}.
By integrating the power over these round trips, the energy efficiency can be calculated.
Some studies apply this method to partial cycles, each at a constant C-rate, thereby enabling the calculation of the $SoC$-dependency of $\eta_{RT,e}$ \cite{Liu2017}.
In other cases, the round-trip efficiency is split into charge and discharge energy efficiency, by using knowledge of the cell's electromotive force, or open-circuit voltage \cite{Kang2012a,Bobanac2021}.
However, few studies provide energy efficiency results based on drive-cycle data and there are only few papers that study the energy-efficiency fade \cite{Ahmadi2014}.

This paper presents an algorithm to determine the round-trip efficiency $\eta_{RT,e}$ and its fade thereby making three main contributions:
\begin{enumerate}
\item An algorithm to determine $\eta_{RT,e}$ based on drive cycles.
\item A methodology to compare the $\eta_{RT,e}$ as measured under various conditions, thus quantifying the fade of $\eta_{RT,e}$ over the years.
\item Demonstration of this method on real-world data from battery-electric city buses, over a 3.5-year period.
\end{enumerate}
These contributions are expected to be required for the future implementation of the battery passport.

The contents of the remainder of this paper are organized as follows.
The methodology that determines the round-trip efficiency is described in Section~\ref{sec:methodology}, including a method to compare multiple of these calculated efficiencies over time.
The method is applied to a real-world dataset and discussed in Section~\ref{sec:Results}.
Lastly, the conclusions and future work are presented in Section~\ref{sec:conclusions}.

\section{Round-Trip Efficiency Estimation}\label{sec:methodology}
Given data sampled {\color{changed2} at a total of} $N$ discrete time steps with sampling time $\Delta{}t$, such that $t = t_0,t_0+\Delta{}t,...,t_{N-1}$, the round-trip energy efficiency can be calculated as
\begin{equation}\label{eq:RTefficiency}
\begin{aligned}
\eta_{RT,e} =& {\color{changed} E_{dis}/E_{chg}} \\
=& \frac{ \sum_{\tau  = t_{\color{changed2}start}}^{t_{\color{changed2}end}} \left.-I(\tau) U_t(\tau)\right|_{dis}}{\sum_{\tau = t_{\color{changed2}start}}^{t_{\color{changed2}end}} \left.I(\tau) U_t(\tau)\right|_{chg}} = \frac{\sum_{\tau \in t_{dis}}-I(\tau) U_t(\tau) }{ \sum_{\tau \in t_{chg}} I(\tau) U_t(\tau)},
\end{aligned}
\end{equation}
where $I(t)$ is the battery current, $U_t(t)$ is the battery terminal voltage, and the subscripts $dis$ and $chg$ indicate the discharge ($I(t)<0$) and charge ($I(t)>0$) situation, respectively.
The time instances $t_{\color{changed2}start}$ and $t_{\color{changed2}end}$ define the beginning and end of the round trip, which are characterized by corresponding $SoC$ values:
\begin{equation}\label{eq:SOCdef}
SoC(t_{\color{changed2}end}) := SoC(t_{\color{changed2}start})\;,
\end{equation}
where $SoC$ is calculated as
\begin{equation}
\textstyle SoC(t) = SoC(t_0) + \sum_{t = t_0}^{t_{N-1}} I(t)/C_n\;,
\end{equation}
where $C_n$ is the nominal battery capacity.
Because the round trips are relatively short, this coulomb-counting method is assumed to be sufficiently accurate.
Furthermore, $t_{dis}$ and $t_{chg}$ are
\begin{equation}
\begin{aligned}
t_{dis} &:= \left\{\, t_{dis} \in t \,|\, t_{\color{changed2}start} \leq t_{dis} \leq t_{\color{changed2}end} \wedge I(t_{dis})< 0 \,\right\}\\
t_{chg} &:= \left\{\, t_{chg} \in t \,|\, t_{\color{changed2}start} \leq t_{chg} \leq t_{\color{changed2}end} \wedge I(t_{chg})> 0 \,\right\} \;.
\end{aligned}
\end{equation}

To apply \eqref{eq:RTefficiency}, the values for $t_{\color{changed2}start}$ and $t_{\color{changed2}end}$ are to be determined.
Even though any two time instances that fulfill \eqref{eq:SOCdef} could be chosen, additional requirements are formulated.
Firstly, to minimize the effect of any polarization voltage on $U_t(t_{\color{changed2}start})$, $t_{\color{changed2}start}$ is required to be after a period of low current;
{\color{changed}
\begin{equation}\label{eq:conditiont1}
t_{\color{changed2}start} = \{t_{\color{changed2}start} \in t | \left( \exists t_{r}\right)\left[I(t_{\color{changed2}start}-t_{r}) < I_{r} \forall t_{r} < t_{r,min}\right]\}.
\end{equation}
This rest period is defined by a minimum duration $t_{r,min}$ and a maximum current $I_{r}$, which is small enough for the battery to be approximately in rest.
If $I(t) < I_{r}$ for at least $t_{r,min}$ a point is defined as a possible start of a round trip.

}
Secondly, the end of the round trip is marked by $t_{\color{changed2}end}$ which according to \eqref{eq:SOCdef} has the same $SoC$ as $t_{\color{changed2}start}$.
{\color{changed} Additional constraints are added to ensure a minimum duration $t_{min}$ and maximum duration $t_{max}$ of the round-trip:
\begin{equation}\label{eq:conditiont2}
t_{\color{changed2}end} = \{\; t_{\color{changed2}end} \in t \;|\; \;\eqref{eq:SOCdef}\; \;\wedge\; t_{min} < (t_{\color{changed2}end}-t_{\color{changed2}start}) < t_{max} \;\}\;.
\end{equation}
Whereas in theory \eqref{eq:SOCdef} is considered to be fulfilled if $SoC(t_{\color{changed2}start})$ and $SoC(t_{\color{changed2}end})$ are equal, in practise a small $\Delta{}SoC$ between the two values is allowed.}
{\color{changed2} In case multiple consecutive data points fall within the range $SoC(t_{\color{changed2}start})\pm\Delta{}SoC$, the middle of these points is selected to represent $t_{end}$}.

\subsection{Energy Efficiency and Uncertainty}
Once a round trip has been defined by $t_{\color{changed2}start}$ and $t_{\color{changed2}end}$, $\eta_{RT,e}$ can be calculated through numerical integration of the measured signals $I(t)$ and $U(t)$ by applying \eqref{eq:RTefficiency}.
Additionally, it is of interest to determine the standard error of the efficiency values through error propagation.
For this purpose, the round-trip efficiency is written as $\eta_{RT,e} = E_{dis}/E_{chg}$ {\color{changed2}\eqref{eq:RTefficiency}}, where $E_{dis}$ is the discharged energy and $E_{chg}$ is the charged energy during the round trip.
By using $\frac{\partial\eta_{RT,e}}{\partial E_{dis}} = \frac{1}{E_{chg}}$ and $\frac{\partial\eta_{RT,e}}{\partial E_{chg}} = \frac{-E_{dis}}{E_{chg}^2}$, the standard error of $S_{\eta_{RT,e}}$ can be expressed as
\begin{equation}\label{eq:numintUncert}
\begin{aligned}
S_{\eta_{RT,e}} = \sqrt{ \left(\frac{1}{E_{chg}} S_{E_{dis}}\right)^2 + \left( \frac{-E_{dis}}{E_{chg}^2} S_{E_{chg}}\right)^2}\;,
\end{aligned}
\end{equation}
where $S_{E_{dis}}$ and $S_{E_{chg}}$ are the standard errors of the respective energies.
Both $E_{chg}$ and $E_{dis}$ are the result of numerical integration of the battery power. Therefore, these standard errors can be written as
\begin{equation}
S_{E_{dis}}(t_{dis}) = \sqrt{  \sum_{\tau \in t_{dis}} \left(\,U_t(\tau)^2S_{I}(\tau)^2 +  I(\tau)^2S_{U_t}(\tau)^2\,\right)\Delta{}t }\,
\end{equation}
and likewise for $S_{E_{chg}}(t_{chg})$.
In the above equations, the standard error for the voltage and current measurements, respectively $S_{U_t}(t)$ and $S_{I}(t)$ represent the systematic uncertainty due to sensor nonlinearity, offset, and resolution, as based on typical sensor specifications, and the observed random uncertainty due to noise.

\begin{figure*}[!t]
  \centering
  \includegraphics[width=\textwidth,trim= 0 0.5cm 0 0]{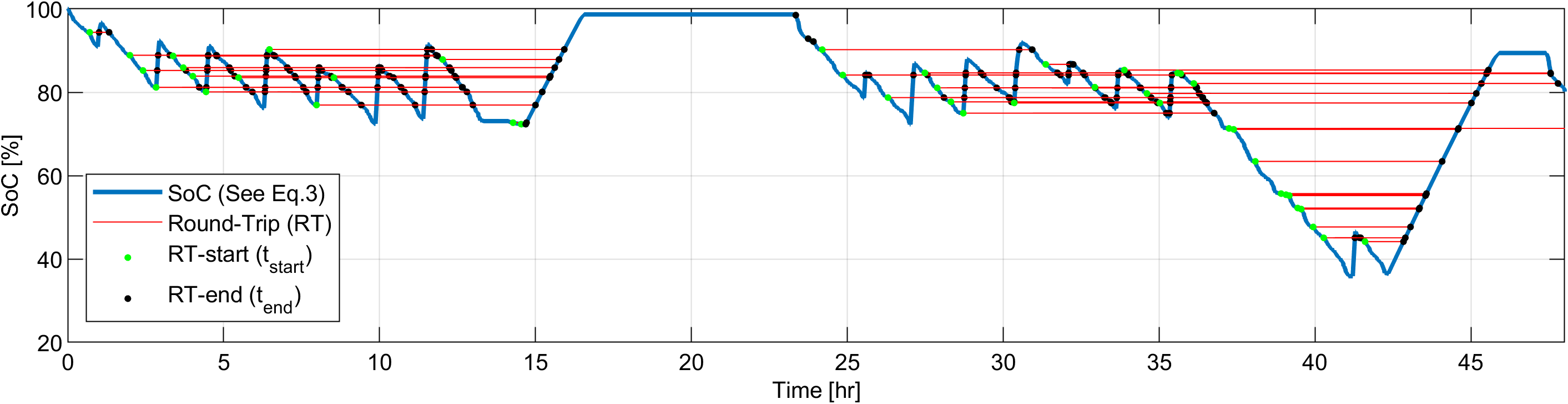}
  \caption{The $SoC$ of Vehicle A in October 2019 during {\color{changed} 48\,hours}. The red lines indicate the identified round trips, {\color{changed} the start and end of which are marked by the time instances $t_{\color{changed2}start}$, respectively $t_{\color{changed2}end}$, as defined by \eqref{eq:conditiont1} and \eqref{eq:conditiont2}. The $SoC$ decrease {\color{changed2} between $t = 36$\,hours and $t = 41$\,hours} is due to a period of driving without intermittent charging events. {\color{changed2}This initiates several shorter round-trips at lower $SoC$ values, which are ended during the subsequent charging session between $t = 42$\,hours and $t = 46$\,hours.}}}\label{fig:roundtrips}
\end{figure*}

\subsection{Theoretical Relation to Impedance}\label{sec:Thevenin}
The energy efficiency can be related to the battery impedance by considering the Th\'evenin model.
This model considers the terminal voltage ${U_t}(t)$ to be a sum of the electromotive force $U_{EMF}(SoC)$ and the voltage drop over an ohmic resistance $R_0$:
\begin{equation}\label{eq:thevenin}
U_t(t) = U_{EMF}(SoC) + R_0 I(t)\,.
\end{equation}
By applying \eqref{eq:RTefficiency} and assuming constant currents $I_{dis} = -I_{chg} = I_c$ and constant $U_{EMF}(SoC) = U_{EMF,c}$ the efficiency can be approximated as
\begin{equation}
\eta_{RT,e} \approx \frac{U_{EMF,c} - R_0 I_c}{U_{EMF,c} + R_0 I_c}\,.
\end{equation}
By further assuming $I_c << U_{EMF,c}/R_0$ this reduces to
\begin{equation}\label{eq:theoryImpedance}
\eta_{RT,e} \approx 1-\frac{2R_0I_c}{U_{EMF,c}}\;.
\end{equation}
This relation indicates that the efficiency is dictated by the impedance $R_0$ and decreases linearly with increasing currents $I_c$.
Since the impedance is both temperature and $SoC$ dependent, and $U_{EMF}$ is also $SoC$ dependent, the efficiency is also a function of these conditions.

Assuming $U_{EMF}(SoC)$ does not change over the batter{\color{changed}y} lifetime, the expected efficiency fade with respect to the Beginning of Life (BoL) situation can be quantified as
\begin{equation}\label{eq:theoryExpectedFade}
\left.\eta_{RT,e}\right|_{BoL}- \eta_{RT,e} \approx \frac{2(R_0-R_{0,BoL})I_c}{U_{EMF,c}}\;,
\end{equation}
which shows that the expected decrease scales linearly with any increase in internal resistance.

\subsection{Quantifying Conditions}\label{sec:ConditionDefinitions}
Because $\eta_{RT,e}$ is expected to be dependent on several conditions, such as $SoC$, $DoD$, C-rate, and $T$, the following four conditions are defined for each round-trip:
\begin{equation}\label{eq:conditionSoC}
SoC_{RT} = \textstyle 1/n\sum_{t = t_{\color{changed2}start}}^{t_{\color{changed2}end}} SoC(t)
\end{equation}
\begin{equation}\label{eq:conditionDoD}
DoD_{RT} = \textstyle \max(SoC(t_{RT}))-\min(SoC(t_{RT}))
\end{equation}
\begin{equation}\label{eq:conditionCrate}
{\color{changed}C_1 = }\, \text{RMS C-rate}_{RT} = \textstyle \sqrt{1/n\sum_{t = t_{\color{changed2}start}}^{t_{\color{changed2}end}} (I(t)/C_n)^2 }
\end{equation}
\begin{equation}\label{eq:conditionTemp}
{\color{changed}C_2 = }\, T_{RT} = \textstyle 1/n\sum_{t = t_{\color{changed2}start}}^{t_{\color{changed2}end}} T(t)\;,
\end{equation}
where $n$ is the total number of measurement points in the round trip and $t_{RT} = t_{\color{changed2}start},t_{\color{changed2}start}+\Delta{}t,...,t_{\color{changed2}end}$, with $t_{\color{changed2}end} = t_{\color{changed2}start}+n\Delta{}t$.
In Section~\ref{sec:ComparingConditions} these conditions will be compared and ranked based on measurement data. For the remainder of this section, it is assumed that there are only two conditions that are statistically relevant, as will be confirmed in Section~\ref{sec:ComparingConditions}.

\subsection{Efficiency Maps and Efficiency Fade}\label{sec:efficiencyFade}
For every round trip $i$ {\color{changed2} of the in total $n_{RT}$ round trips,} the energy efficiency $\eta_{RT,e,i}$ can be calculated, together with the corresponding conditions $C_{1,i}$ and $C_{2,i}$, which are a subset of the conditions described above.
The challenge now lies in comparing these efficiency values determined under different conditions.
To this end, a linear regression model is sought that provides an estimate of the round-trip efficiency $\hat{\eta}_{RT,e}$ as function of the two conditions:
\begin{equation}\label{eq:linearregression}
\hat{\eta}_{RT,e} = \beta_1 C_1 + \beta_2 C_2 + \beta_3\;,
\end{equation}
where $\beta_1$, $\beta_2$, and $\beta_3$ are the model parameters.
In this case, weighted multiple linear regression \cite[p. 466]{montgomery2006} is applied with weights $w_i = S_{{\eta}_{RT,e,i}}^{-2}$, which are based on the standard error defined in \eqref{eq:numintUncert}.
{\color{changed} The solution to this least-squares problem can be found analytically by solving

\begin{equation}
\left(\mathbf{X}'\mathbf{WX}\right)\underline{\beta} = \mathbf{X}'\mathbf{W}\underline{y}
\end{equation}
for $\underline{\beta}$ where
\begin{equation}\nonumber
\begin{aligned}
\mathbf{X} =& \begin{bmatrix}
  \begin{bmatrix}
    C_{1,1} \\
    C_{1,i} \\
   \vdots     \\
    C_{1,{\color{changed2}n_{RT}}}
  \end{bmatrix} & \hspace{-0.2cm}\begin{bmatrix}
    C_{2,1} \\
    C_{2,i} \\
   \vdots     \\
    C_{2,{\color{changed2}n_{RT}}}
  \end{bmatrix} & \hspace{-0.2cm}\begin{bmatrix}
    1 \\
    1 \\
   \vdots  \\
    1
  \end{bmatrix}
\end{bmatrix}, \mathbf{W} = \text{diag}\begin{pmatrix}
    w_{1} \\
    w_{i} \\
   \vdots     \\
    w_{{\color{changed2}n_{RT}}}
  \end{pmatrix},\\
\underline{\beta} =& \begin{bmatrix}
 \beta_1 \\ \beta_2 \\ \beta_3 \end{bmatrix},\;\;\;\; \text{and}\;\;\;\; \underline{y} = \begin{bmatrix} \eta_{RT,e,1} \\ \eta_{RT,e,{\color{changed2}i}} \\ \vdots     \\ \eta_{RT,e,{\color{changed2}n_{RT}}} \end{bmatrix}.
\end{aligned}
\end{equation}
}

In addition to estimates for the parameters $\beta_1$, $\beta_2$, and $\beta_3$, the weighted least-squares model {\color{changed} allows for calculation of the} standard errors for these parameters and for prediction of $\hat{\eta}_{RT,e}$ at conditions $(C_1, C_2)$, even if that combination of conditions was not measured during any of the round trips.
Furthermore, when multiple datasets are available, several linear regression models $\hat{\eta}_{RT,e}(C_1, C_2)$ can be determined.
Applying this method to sets of data gathered at different moments in time enables the possibility to compare $\hat{\eta}_{RT,e}$ through time under the same conditions.

\section{Results and Discussion}\label{sec:Results}
The method described in Section~\ref{sec:methodology} is applied to a real-world dataset originating from a fleet of Battery-Electric Buses (BEBs).
The dataset contains battery pack current $I$, battery pack voltage $U_t$, and average pack temperature $T$ at a sampling frequency of 1\,Hz as reported by the Battery Management System (BMS).
Data is available from three vehicles of the same type; Vehicle A, B, and C.
The dataset spans a time range from August 2019 until December 2022, where only the first 5 days of every month are available.
This data is parsed into 41 separate datasets per vehicle; one per month.

\subsection{Round-Trips and Efficiency Values}\label{sec:RTEvalues}
To exemplify the method described in Section\,\ref{sec:methodology}, data from a single month is analyzed.
Several round trips can be found, based on the signals $I(t)$ and $SoC(I(t))$, as shown in Fig.~\ref{fig:roundtrips}.
The results show the $SoC$ as function of time for a duration of {\color{changed} 48\,hours}.
It can be seen that the battery experiences multiple charging events during the day and that there are prolonged periods of no use during the night.
Based on these 48 hours worth of data, {\color{changed2}241} round trips are identified, as indicated by the horizontal red lines.
These round trips are of different durations and happen at different SoC values.

\subsection{Establishing Relevant Conditions}\label{sec:ComparingConditions}
Using the round trips as shown in Fig.~\ref{fig:roundtrips}, the values for the round-trip efficiency can be calculated according to \eqref{eq:RTefficiency}.
These efficiencies are shown in Fig.~\ref{fig:conditions} as function of the four conditions as defined by (\ref{eq:conditionSoC},...,\ref{eq:conditionTemp}).
The Spearman correlation coefficient $\rho$ is used to quantify the correlation between each respective condition and $\eta_{RT,e}$ \cite{Spearman1904}.
\begin{figure}[t!]
  \centering
  \includegraphics[width=\columnwidth]{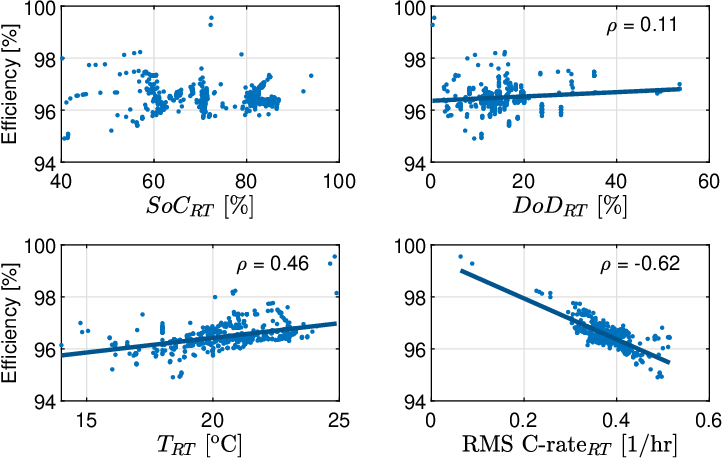}
  \caption{Round-trip energy efficiency {\color{changed2}$\eta_{RT}$} of Vehicle A in October 2019 as function of four different conditions defined by {\color{changed2}(\ref{eq:conditionSoC},...,\ref{eq:conditionTemp})}. A trend line is visualized in combination with the Spearman correlation coefficient $\rho$. {\color{changed} No trend line is visualized in the top-left figure, because no statistically relevant correlation could be found.}}\label{fig:conditions}
\end{figure}

First of all, the results generally indicate an average $\eta_{RT,e}$ of approximately 96.2\%.
When investigating the correlations, it becomes clear that for $SoC_{RT}$ and $DoD_{RT}$, it is either not possible to establish a significant statistical relation, {\color{changed} as is the case for $SoC_{RT}$}, or the correlation is relatively low ($\rho = 0.11$).
{\color{changed} The fact that $\eta_{RT,e}$ does not seem to depend on $SoC_{RT}$ is in contrast to the general consensus that $U_{EMF}$ and in reality also $R_0$, as defined in Section\,\ref{sec:Thevenin}, depend on $SoC$.
The fact that this correlation is not found in the data, could be a consequence of the specific drive cyles found in the {\color{changed2}BEB} application.
Large discharges of the battery seldomly happen, causing $SoC_{RT}$ to vary mostly between 60\% and 80\%, where the influence of this parameter on $\eta_{RT,e}$ is limited.
However, this observation does indicate that the relevant conditions might be application-dependent.

More significant correlations are found in $T_{RT}$, which shows} a slightly larger positive correlation ($\rho = 0.46$).
This confirms that $\eta_{RT,e}$ becomes larger with increasing temperature.
Lastly, there is a strong negative correlation between the RMS C-rate and $\eta_{RT,e}$, which is also expected based on \eqref{eq:theoryImpedance}.
Based on these results, only $T_{RT}$ and RMS C-rate$_{RT}$ are considered influential conditions for the remainder of this study.
Therefore, \mbox{$C_1 = \,$RMS C-rate$_{RT}$} and $C_2 = T_{RT}$.

\subsection{Resulting Efficiency Maps}
Using the theory detailed in Section\,\ref{sec:efficiencyFade} the efficiency values can be visualized as function of these main conditions $T_{RT}$ and RMS C-rate$_{RT}$.
A linear regression model as described in \eqref{eq:linearregression} is based on this data, and visualized in Fig.~\ref{fig:regression}.
\begin{figure}[t!]
  \centering
  \includegraphics[width=\columnwidth,trim= 0 0.3cm 0 0]{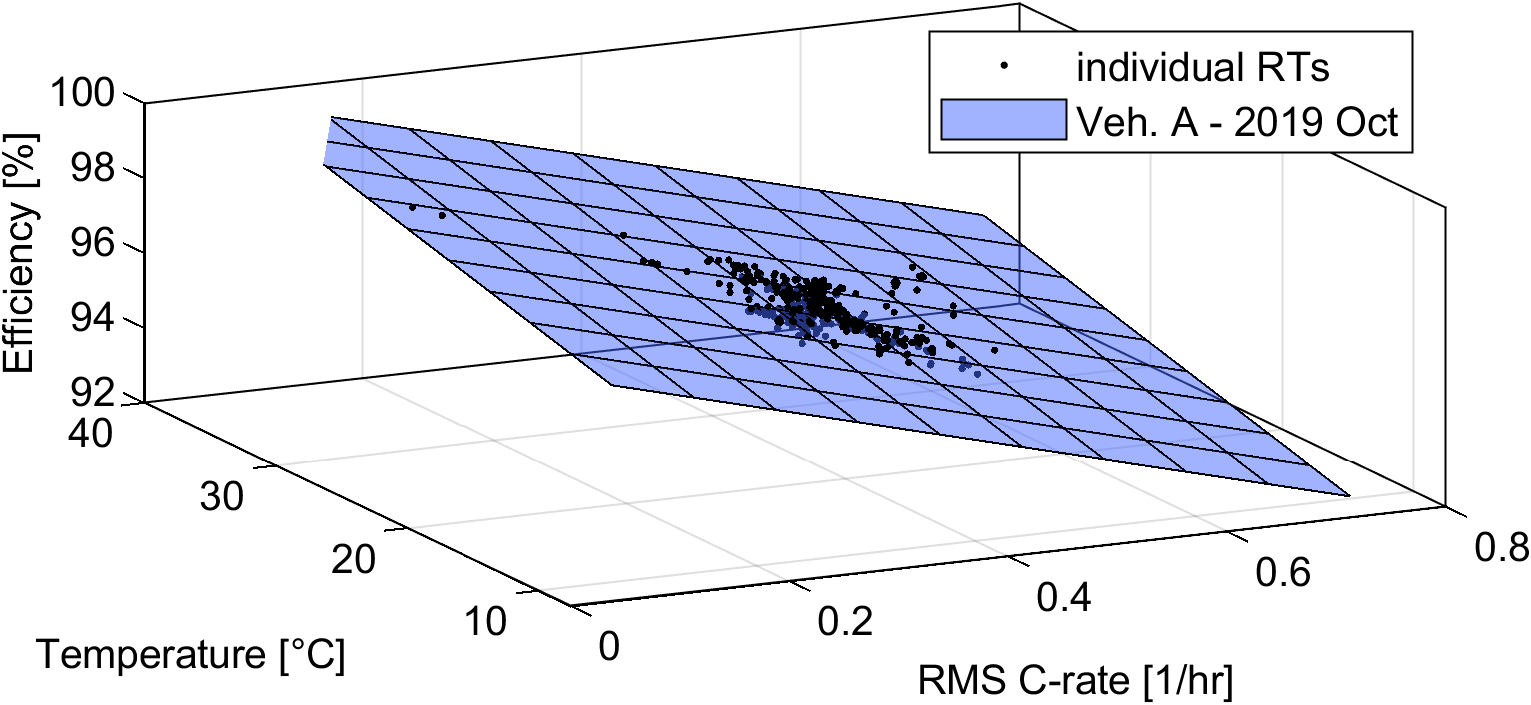}
  \caption{{\color{changed2} The round-trip efficiency} ${\eta}_{RT,e}$ as function of temperature $T_{RT}$ and RMS C-rate$_{RT}$ for Vehicle A in October 2019, visualized as points. The {\color{changed2}plane representing the estimated efficiency} $\hat{\eta}_{RT,e}$ is also shown.}\label{fig:regression}
\end{figure}
%
The estimated parameters of this model are listed in Table\,\ref{tab:regressionparameters}.
Values and standard errors for all three model parameters are provided and indicate once again a negative correlation between the efficiency and RMS C-rate$_{RT}$, and a slight positive correlation between $T_{RT}$ and efficiency.
The low p-values indicate that none of the three model parameters are likely to be zero, confirming that the correlations with $T_{RT}$ and RMS C-rate$_{RT}$ are significant.
Nevertheless, the Adjusted R$^2$ value of the regression model is 0.5, indicating that the model does not fully describe the variance encountered in the efficiency values and could be improved in future work.

\begin{table}[t!]
\centering
\caption{Weighted multiple linear regression results for Vehicle A in Oct 2019.}\label{tab:regressionparameters}
\begin{tabular}{|c|c|c|c|}
  \hline
  \bfseries Parameter & \bfseries Value & \bfseries Standard Error & \bfseries p-value\\
  \hline
  $\beta_1$ & -7.94 [\%$\cdot{}$hr] & 0.33 [\%$\cdot{}$hr]  & $2\cdot10^{-89}$\\
  $\beta_2$ & 0.084 [\%/$^{\circ}$C]       & 0.005 [\%/$^{\circ}$C] & $8\cdot10^{-45}$\\
  $\beta_3$ & 97.76 [\%]            & 0.16 [\%]  & 0               \\
  \hline
\end{tabular}
\end{table}

\subsection{Efficiency Fade}
The procedure to find a linear regression model, based on several days of data, can be repeated for multiple parts of the dataset.
In this case, the first 5 days of every month are used per vehicle.
For each of these 41 months, a linear regression model $\hat{\eta}_{RT,e}($RMS C-rate$_{RT},T_{RT})$ is determined.
Several of these models are visualized in Fig.~\ref{fig:regresssionMultiple} for a single vehicle.
The figure shows that the different models are generally in agreement, {\color{changed} because they approximately span the same plane}.
Some regression models predict efficiencies exceeding 100\% for high temperatures and low C-rates, {\color{changed} as indicated by the clipped corners,} which is a consequence of the linear extrapolation.
\begin{figure}[t!]
  \centering
  \includegraphics[width=\columnwidth,trim= 0 0.5cm 0 0]{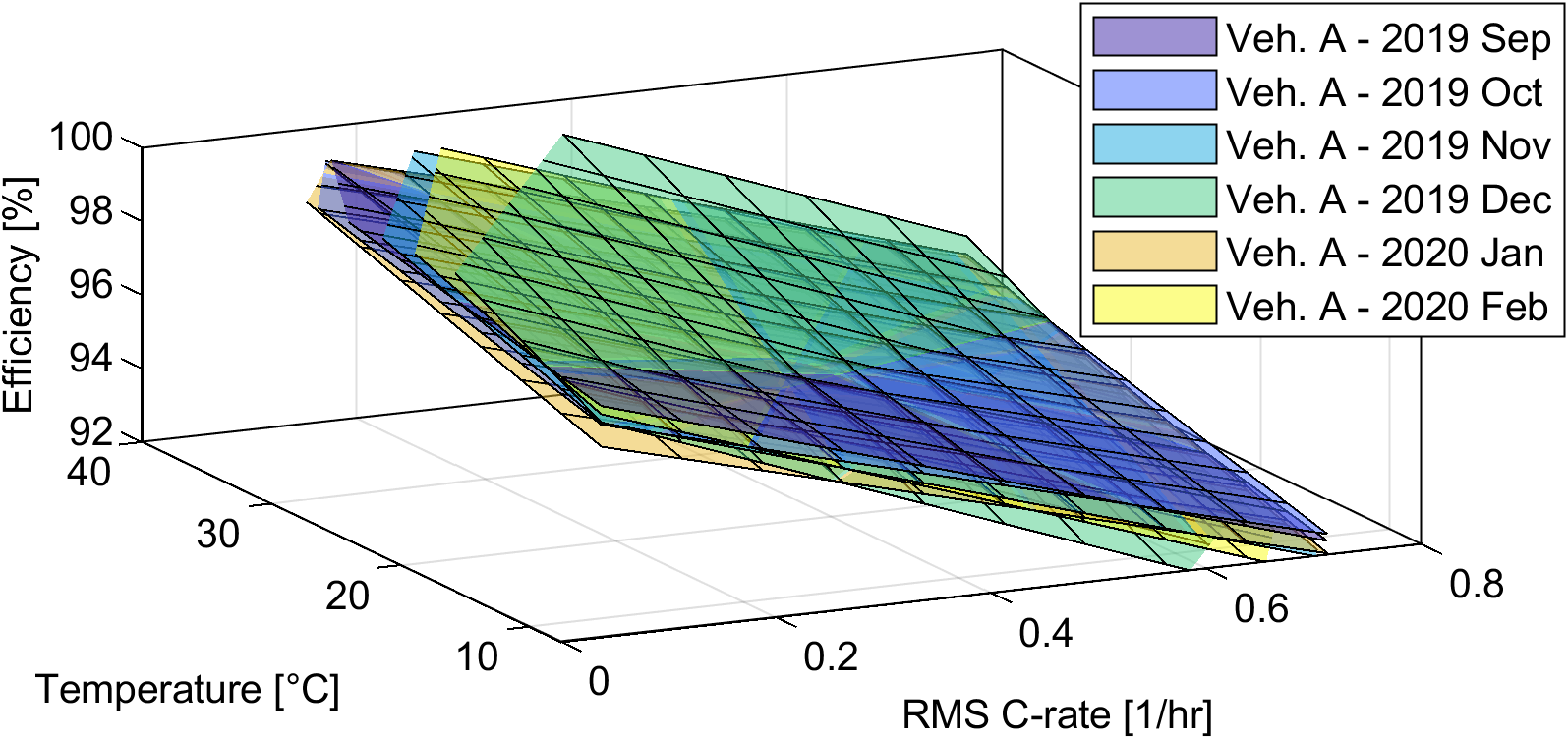}
  \caption{{\color{changed2} The estimated round-trip energy efficiency} $\hat{\eta}_{RT,e}$ as function of {\color{changed2} temperature} $T_{RT}$ and RMS C-rate$_{RT}$ for Vehicle A at different moments in time.}\label{fig:regresssionMultiple}
\end{figure}

The linear regression models are all evaluated at a particular condition combination, to allow for a comparison of $\hat{\eta}_{RT,e}$ through time at the same conditions.
In this case, specific conditions $C_1^*$ = C-rate$_{RT}^*$
and $C_2^* = T_{RT}^*$
are chosen such that these represent the averages of the entire dataset.
Figure~\ref{fig:fade} shows $\hat{\eta}_{RT,e}(C_1^*, C_2^*)$ as function of time.

The results of Vehicle A show that $\hat{\eta}_{RT,e}$ generally decreases as function of time, although not monotonically.
Occasional increases in energy efficiency are plausible because of relaxation of long-term build up of polarization voltages, which is not always related to irreversible aging processes.
The trend line shows yearly oscillations, which could be an indication that there is still a temperature dependency in the results.
Altering the model structure of \eqref{eq:linearregression} to better fit the effect of temperature might improve this.
The results for vehicle A indicate that $\hat{\eta}_{RT,e}$ reduces from 97.45\% to 96.59\%{\color{changed2}, thus by 0.86\,p.p.,} over the 3.5-year period.

When considering the results from vehicle B, it becomes clear that a significant efficiency reduction cannot always be concluded.
Large uncertainties in individual estimates of $\hat{\eta}_{RT,e}$ can be seen, often caused by reduced use of the vehicle, causing a fewer number of round trips to be detected.
The fact that no decrease in efficiency is observed could be caused by the intermediate replacement of battery modules on the vehicle.
Lastly, vehicle C, similar to vehicle A, shows a non-monotonically decreasing energy efficiency.

When summarizing the results, as shown in Table~\ref{tab:finalresults}, it becomes clear that an energy-efficiency fade of 0.46\,percent point (p.p.) is observed on average for these three vehicles.
This seemingly minor decrease can be connected to the impedance through \eqref{eq:theoryExpectedFade}, which indicates that an efficiency decrease from approximately 97\% to a 96.5\%, corresponds to a relative impedance increase of 17\%, which is plausible.
\begin{figure}[t!]
  \centering
  \includegraphics[width=\columnwidth,trim= 0 0.3cm 0 0]{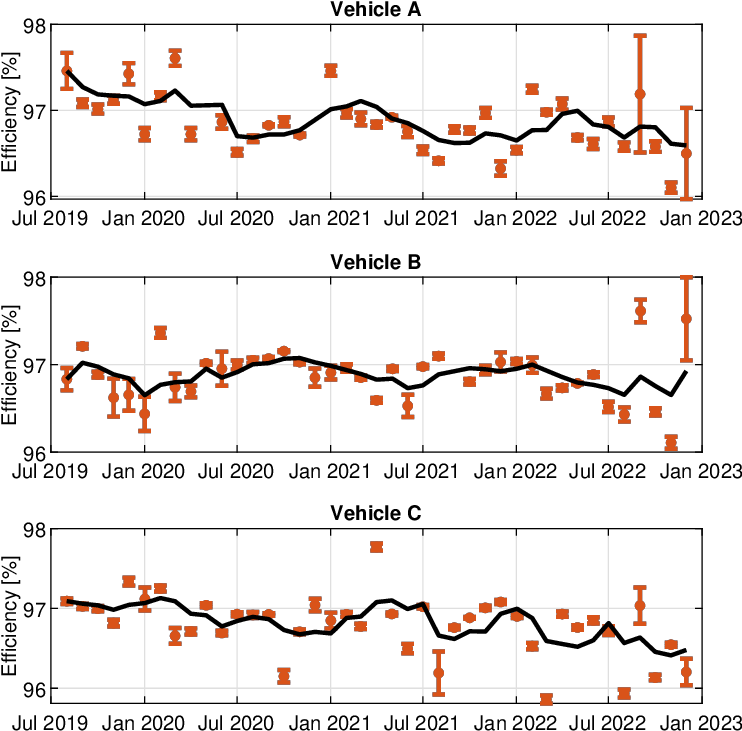}
  \caption{{\color{changed2} The estimated round-trip energy efficiency} $\hat{\eta}_{RT,e}$ at C-rate$_{RT}^*$ and $T_{RT}^*$ for all three vehicles as function of time. The error bars indicate the 95\% confidence bounds. The black lines indicate the three-month moving average.}\label{fig:fade}
\end{figure}

\subsection{Implications for Battery Passport}

The presented results shed some light on the implications of the inclusion of an efficiency-fade quantification in the battery passport.
As shown by the results, and also based on the theoretical relation between impedance and energy efficiency, only a slight decrease in energy efficiency is to be expected over the battery lifetime.
Accurately quantifying this fade can be challenging, also considering the fact that the momentary energy efficiency depends on several conditions, which should be accounted for in the efficiency-fade calculation.
Based on the results, it is established that the efficiency will generally decrease, yet momentary increases are also possible, which should also be accounted for in the battery-passport implementation.
The energy-efficiency fade quantifies the combination of several physical degradation effects in an interpretable way, yet is not always trivial to calculate.

\section{Conclusions and Future Work}\label{sec:conclusions}
This paper presents an algorithm to determine the round-trip energy efficiency and energy-efficiency fade based on measured data from a battery pack.
The algorithm detects events that mark the beginning and end of round trips and calculates the energy efficiency through numerical integration.
The method is demonstrated on data from several battery-electric buses and shows that the average energy efficiency is 96\%.
Of the four conditions found in literature, the energy efficiency is shown to be most strongly correlated to RMS C-rate and battery temperature.
These two conditions are then used as the dependent variables in a linear regression model.
Multiple of these linear regression models are compared over 41 months.
The results show a battery degradation up to 0.86\,p.p. in the course of 3.5\,years.

\begin{table}[t!]
\centering
\caption{{\color{changed2}The estimated round-trip efficiency} $\hat{\eta}_{RT,e}$ and its fade at {\color{changed2} the specific conditions} C-rate$_{RT}^*$ and $T_{RT}^*$ for three vehicles.}\label{tab:finalresults}
\begin{tabular}{|c|c|c|c|}
  \hline
  \bfseries Vehicle & $\hat{\eta}_{RT,e}$ Aug.'19 & $\hat{\eta}_{RT,e}$ Dec.'22 & $\hat{\eta}_{RT,e}$-fade \\
  \hline
  A & 97.45\,\% & 96.59\,\% &  0.86\,p.p.\\
  B & 96.83\,\% & 96.92\,\% & -0.09\,p.p.\\
  C & 97.09\,\% & 96.48\,\% &  0.61\,p.p.\\\hline
  Average & 97.12\,\% & 96.66\,\% &  0.46\,p.p.\\
  \hline
\end{tabular}
\end{table}

Although the method presented in this paper achieves quantification of the energy-efficiency fade over time, several improvements could be considered.
Firstly, the algorithm depends on several parameters.
A sensitivity study could improve the accuracy of the result by changing these values.
{\color{changed} Furthermore, the results of the algorithm might depend heavily on the application in which the battery is used.
Different drive cycles will influence the number of round trips that can be identified and might result in different combinations of relevant conditions.
A Battery Energy Storage System (BESS), as used for grid peak-shaving, is an interesting future case to study, because the round-trip energy efficiency has a direct relation to the Total Cost of Ownership (TCO) of this system.
Therefore, any energy efficiency decrease might influence the economic lifetime of such a BESS.
However, based on the results presented here, which are of an EV application, the energy efficiency appears to change minimally over a multiple years.}
Thirdly, the regression model could be improved to better describe the temperature dependency of the round-trip efficiency.
Lastly, to make the algorithm useful for a battery passport, an online implementation would be required.

\bibliographystyle{./IEEEtranBST2/IEEEtran}
\IEEEtriggeratref{2}
\bibliography{./IEEEtranBST2/IEEEabrv,./Publications-VPPC23_corrected_IEEE}

\end{document}